\documentclass[prl,aps,twocolumn,epsfig]{revtex4}
\usepackage{graphicx}% Include figure files
\usepackage{dcolumn}% Align table columns on decimal point
\usepackage{bm}% bold math
\usepackage{amsmath}
\usepackage{amsfonts}
\usepackage{amssymb}
\usepackage{graphicx}
\usepackage{textcomp}
\usepackage{color}
\usepackage[ngerman, english]{babel}
\makeatletter

\newcommand{\Rmnum}[1]{\expandafter\@slowromancap\romannumeral #1@}
\makeatother
\begin{document}
\title {CdTe-HgTe core-shell nanowire growth controlled by RHEED}
\author{M. Kessel$^1$, J. Hajer$^1$, G. Karczewski$^2$, C. Schumacher$^1$,\\
C. Br\"une$^1$, H. Buhmann$^1$ and L. W. Molenkamp$^1$}
\vspace{1.5cm}
\address{$^1$ Physikalisches Institut (EP3), Universit\"at W\"urzburg, Am Hubland, 97074 W\"urzburg, Germany.}
\address{$^2$ Institute of Physics, PAS, Al. Lotni\'{o}w 32/46, 02-668 Warsaw, Poland.}
%\date{\today}
\begin{abstract}
We present results on the growth of CdTe-HgTe core-shell nanowires, a realization of a quasi one-dimensional heterostructure of the topological insulator HgTe. The growth is a two step process consisting of the growth of single crystalline zinc blende CdTe nanowires with the vapor-liquid-solid method and the overgrowth of these wires with HgTe such that a closed shell is formed around the CdTe core structure. The CdTe wire growth is monitored by RHEED allowing us to infer information on the crystal properties from the electron diffraction pattern. This information is used to find and control the optimal growth temperature. High quality single crystal CdTe nanowires grow with a preferred orientation. For the growth of the conductive HgTe shell structure we find that the supplied Hg:Te ratio is the crucial parameter to facilitate growth on all surface facets.
\end{abstract}

\maketitle

\section{Introduction}
Two-,\textsuperscript{\cite{bernevig2006,koenig2007}} three-\,\textsuperscript{\cite{fu2007,bruene2011}} and quasi one-dimensional\,\textsuperscript{\cite{egger2010,peng2010}} topological insulators (TIs) are a relatively new field in condensed matter physics, and have already attracted much attention. They are very interesting due to the topological protected electronic surface states,\textsuperscript{\cite{fu2007}} which boast many distinct properties such as spin-momentum coupling.\textsuperscript{\cite{hasan2010,moore2010}}\\
An ideal TI hosts conducting surface states, but has an insulating bulk. However, in many material systems, charge transport investigations reveal a significant bulk contribution. Contrary, MBE-grown HgTe layers show high mobility surface states and no bulk conductance both in two- and three-dimensional structures.\textsuperscript{\cite{bruene2011}} Quasi one-dimensional TI nanowires (NWs) are a novel and interesting topic, as this geometry induces periodic boundary conditions for the surface state along the short perimeter of the NW.\textsuperscript{\cite{egger2010}} First investigations on TI NWs were done on Bi$_2$Se$_3$ based systems.\textsuperscript{\cite{peng2010,xiu2011,hong2014}} However, crystalline quality and charge carrier mobility are inferior to HgTe and furthermore Bi$_2$Se$_3$ has a tendency to be unintentionally doped. More generally, the reduction of size increases the surface to volume ratio. This effect strengthens for a tubular TI, making CdTe NWs shelled with the TI HgTe promising candidates to observe phenomena related to the surface states.\\\\
HgTe NWs can not be grown with the vapor-liquid-solid (VLS) method due to the high vapor pressure of Hg. Attempts to do so resulted in segmented HgTe wires with inclusions of elemental Te.\textsuperscript{\cite{haakennaasen2006}} Here we follow an alternative route: We grow CdTe NWs with the VLS method, which are overgrown with HgTe MBE in a second step in order to obtain CdTe-HgTe core-shell NWs.\\
Conventionally, CdTe NWs show a tendency to grow laterally on the substrate. Previous works only reported short vertical CdTe NWs with wurtzite crystal structure.\textsuperscript{\cite{neretina2007}} Other attempts achieved free-standing zinc blende CdTe NWs, but without specified orientation to the substrate.\textsuperscript{\cite{huang2014}} For uniform HgTe shell growth, we need straight, free standing, preferably oriented and single crystalline zinc blende CdTe NWs.\\
Wojtowicz \textit{et al.} showed, that CdTe NWs can be grown on top of ZnTe NWs.\textsuperscript{\cite{wojtowicz2008}} Moreover, preferably [111] oriented ZnTe NWs can be grown on GaAs substrates with an Au-based catalyst, thus paving a way for obtaining oriented CdTe wires.\textsuperscript{\cite{janik2007}} In this work this approach is refined to achieve the very great uniformity of the NW ensemble. The use of a reflection high energy electron diffraction (RHEED) technique to study the solid-liquid phase transition of the catalyst and the nucleation start has been reported earlier for GaAs NWs with electrons penetrating the nano-crystallites.\textsuperscript{\cite{tchernycheva2006}} We apply the same technique for our CdTe and HgTe-CdTe structures. In addition to the spotty features reported in \cite{tchernycheva2006}, we see streaks in the diffraction patterns, when the electron beam is diffracted during reflection at crystalline facets. The narrow temperature limits for uniform growth are controlled by changes in electron diffraction. High crystalline quality and uniformity of the NWs allows to study the atomic periodicity of NW surfaces by RHEED, which are subsequently overgrown with epitaxial HgTe.

\section{Growth}
\begin{figure}[ht] 
	\begin{center}
		\includegraphics[width=8.5cm]{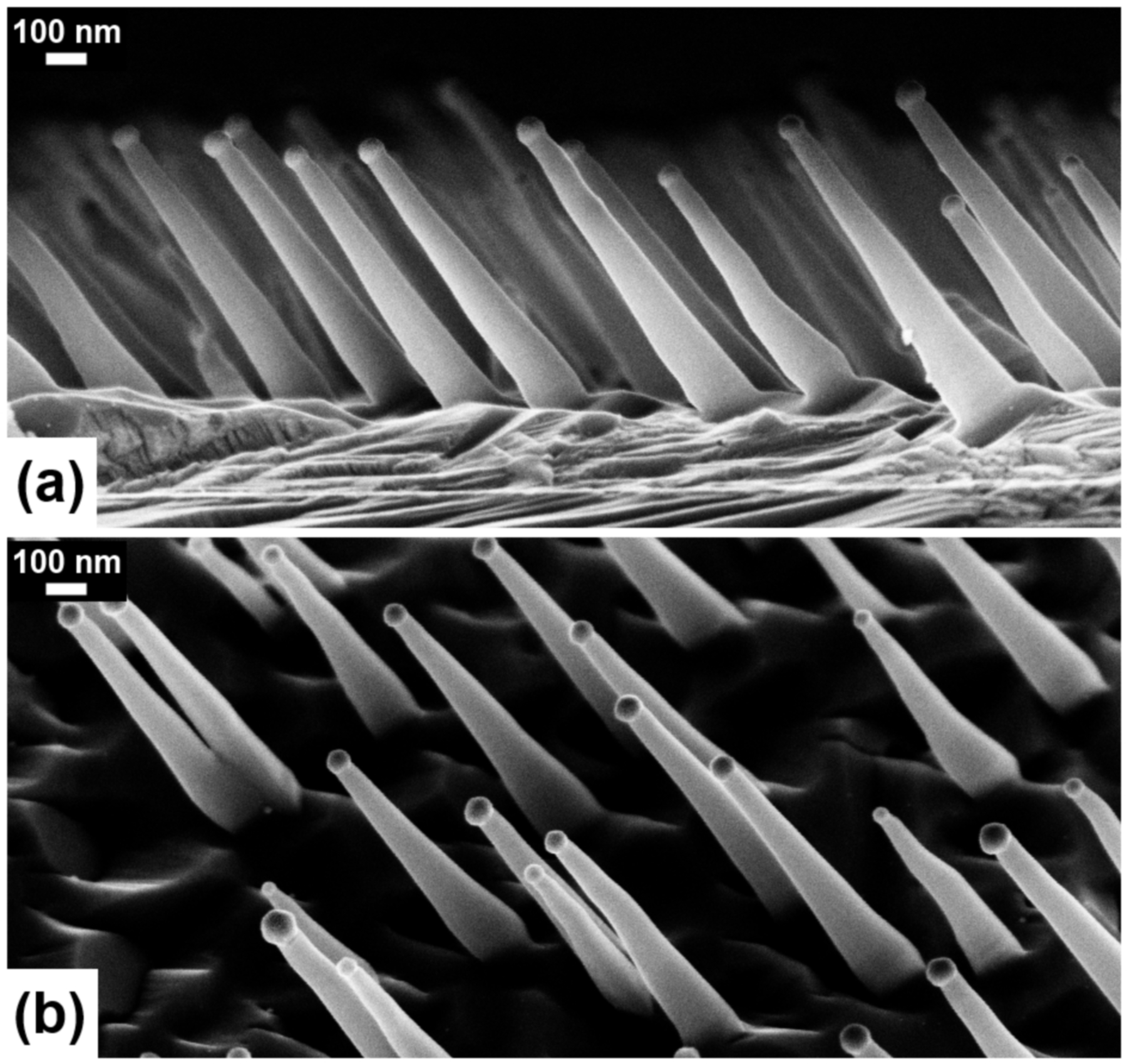}
		\caption[]{SEM pictures with (a) 90$^{\circ}$ and (b) 45$^{\circ}$ view angle of RHEED controlled grown CdTe NWs on a (110) substrate. The substrate temperature was increased by one to two Kelvin per hour for three hours during growth.}
		\label{wire_growth}
	\end{center}
\end{figure}

Starting point of the sample growth are Si-doped GaAs substrates with (110) and (111)B orientation. Our wires are grown in an UHV cluster using a metallization chamber and three separate MBE chambers optimized for the growth of III-V GaAs compounds, wide gap II-VI materials and Hg containing layers, respectively.\\
In order to prepare the substrates, we thermally remove the natural oxide layer followed by the deposition of 1\,nm Au. The NW growth is seeded by liquid eutectic Au-Ga droplets, formed by heating the substrate to 480\,$^{\circ}$C inside the MBE chamber. Subsequently, the NWs are grown with beam equivalent pressures in the range of 5$\cdot$10$^{-7}$ to 1$\cdot$10$^{-6}$\,mbar. Growth starts with a Zn:Te beam pressure ratio of 0.6:1 at 400\,$^{\circ}$C. We find that the growing ZnTe structures push the droplets across the substrate, causing several droplets to merge and become trapped by three-dimensional ZnTe structures. The redistribution and localization of the droplets is complete after about two minutes. We find that these localized droplets seed vertical wire growth.\\
After the localization of the droplets is complete, the Zn supply is replaced by a Cd flux with a Cd:Te ratio of 1.2:1. The growth of the actual CdTe NWs is initiated by carefully lowering the substrate temperature. At 400\,$^{\circ}$C CdTe growth is completely suppressed. We lower the temperature slightly until NW growth starts, while two dimensional CdTe growth remains suppressed. The growth of NWs and the optimal growth temperature can be inferred from the RHEED.\\\\
The optimal substrate temperature to start wire growth is found to be approx.~395\,$^{\circ}$C. The growth rate of CdTe NWs is highly temperature dependent. Growth rates of up to 0.2\,nm/s are observed for substrate temperatures around 395\,$^{\circ}$C. We achieve crystalline NWs with a high aspect ratio. Straight and uniform CdTe NWs grow along [111]B in a temperature range of about 392 - 397\,$^{\circ}$C. When the temperature is too low, the NWs grow randomly along different orientations, leading to pronounced kinks.\\\\
Upon starting NW growth, the sample's radiative loss of energy initially changes significantly, due to the strong surface increase associated with the presence of the NWs. To compensate this effect we increase the substrate temperature slowly by about two Kelvin per hour for three hours during the growth process. Figure~\ref{wire_growth} shows CdTe NWs grown under optimized conditions and the temperature increase is controlled by optimizing the RHEED patterns.\\\\
For long NWs most of the radiated heat remains within the ensemble. In order to grow NWs with lengths up to 2\,\textmu m, a further growth step is necessary where the substrate temperature is held constant at about 400\,$^\circ$C for several hours. The self-organized growth procedure described here results in free-standing CdTe NWs oriented along [111] and a high uniformity, being homogeneous over several cm\textsuperscript{2}.\\\\
In a final step, the NWs are epitaxially overgrown with HgTe at 185\,$^{\circ}$C substrate temperature. Hg is supplied with a beam equivalent pressure in the range of 1$\cdot$10$^{-4}$ to 6$\cdot$10$^{-4}$\,mbar. The Hg:Te ratio can be adjusted using a homebuilt Hg-cell. We find that with a ratio of 200:1 HgTe grows on all NW surfaces. Detailed results can be found in the later sections.

\section{Results and discussion}
\subsection{RHEED studies on CdTe nanowire facets}
\begin{figure*}
	\begin{center}
		\includegraphics[width=18cm]{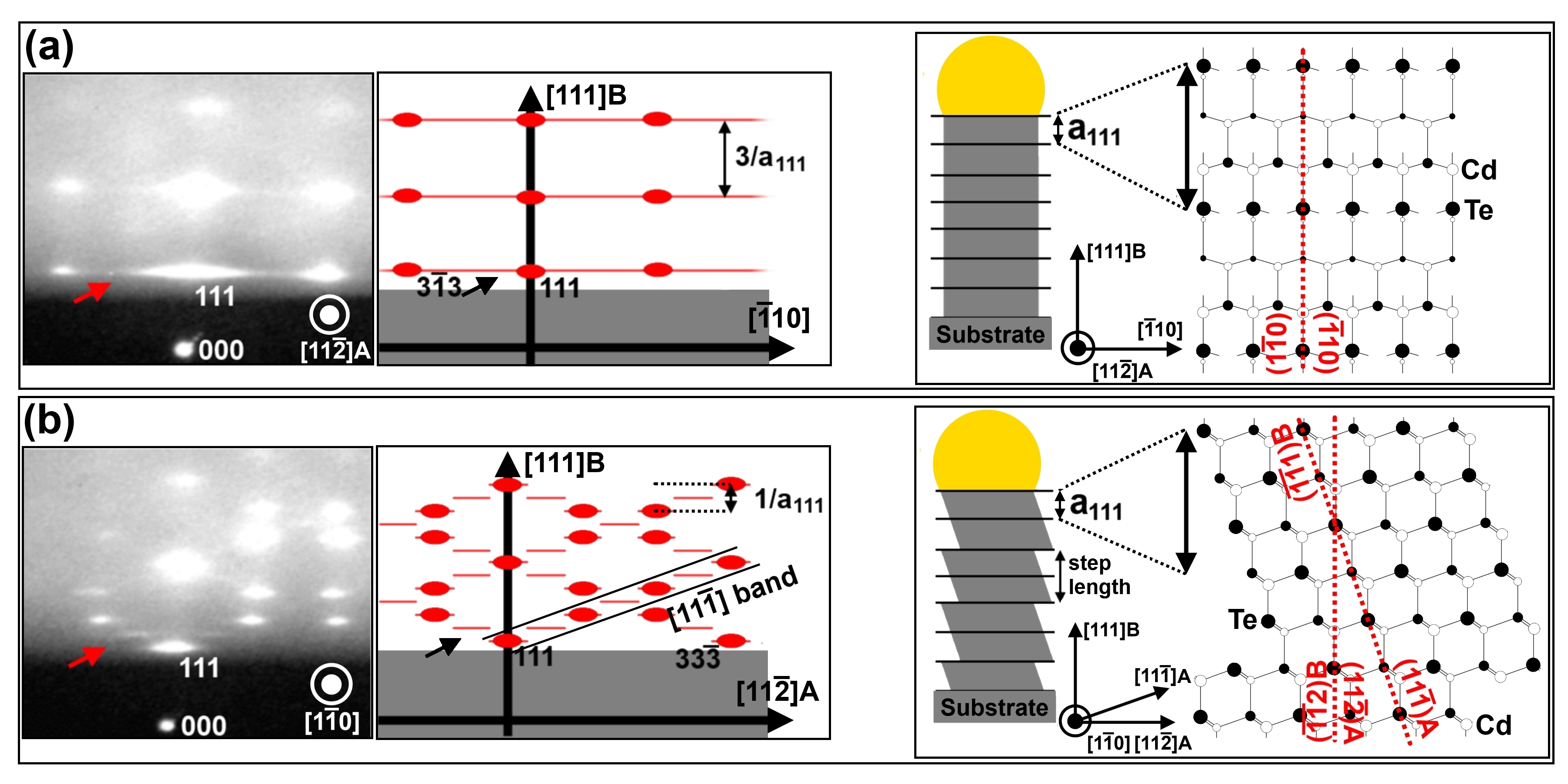}
		\caption[]{RHEED screen, reciprocal space schematic and corresponding real space cross-section looking along (a) [$\overline{1}\overline{1}$2] and (b) [$\overline{1}$10] for a (111)B substrate.}
		\label{RHEED_facets}
	\end{center}
\end{figure*}
In the following section we will discuss how RHEED can be used during growth to gather information on the crystalline structure of the NWs. The method averages over macroscopic sample areas. If the lattice of most of the NWs is identical and furthermore oriented in the same direction, then the interference pattern can be analyzed like the one of a single crystal. Thus, a high ensemble uniformity is needed to allow for detailed analysis and in this case the distance of periodic features in the interference pattern is proportional to the inverse lattice constant. Electrons have a small penetration depth and diffraction during transmission through the bulk region near the surface causes an interference pattern with broadened spots. The period of the spots is a measure for the bulk lattice constant of a single crystal. This is even true for a collection of wires, provided they are all oriented the same.\\\\
In addition to bulk diffraction spots, we observe lines in the RHEED when the NWs are grown long enough to develop sufficiently large side facets of a uniform lattice constant. These lines are caused by diffraction during reflection, and appear perpendicular to the NW side facets. The high uniformity of the NW ensemble allows us to analyze the periodicity of the surfaces by RHEED. Different kinds of crystallographic surface orientations are observed for our CdTe NWs. The sample rotates azimuthally inside the growth chamber. Each NW surface is in the total reflection condition for one azimuth. Twelve different side facets can be identified by comparing the RHEED obtained from different substrate orientations. These are the low index surfaces $\{1\overline{1}0\}$ and $\{11\overline{2}\}$ parallel to the growth axis [111]B. We explain the RHEED pattern of NWs grown vertically on a (111)B substrate. With this geometry one can see two alternating diffraction patterns every 30$^{\circ}$, which are shown in Fig.~\ref{RHEED_facets}.\\\\
The interference pattern obtained by an electron beam along [11$\overline{2}$] as well as its schematic illustration are shown in Fig.~\ref{RHEED_facets}\,(a). The electron beam pointing along [11$\overline{2}$] is reflected from ($\overline{1}$10) facets. Lines perpendicular to this surface emerge in the RHEED. One of these lines is denoted by an arrow in reciprocal space shown left-hand side in Fig.~\ref{RHEED_facets}\,(a). The lines show the same periodicity parallel to the surface as the transmission spots. Therefore the surface atoms show the same period of 1.1\,nm along the [111] direction as in the bulk. Taking into account the atomic positions of the zinc blende unit cell we deduce the cross-sectional view of bulk and ($\overline{1}$10) surface atoms shown right-hand side in Fig.~\ref{RHEED_facets}\,(a).\\\\
The pattern obtained by an electron beam along [1$\overline{1}$0] as well as its schematical illustration are shown in Fig.~\ref{RHEED_facets}\,(b). Zinc blende crystals have a three fold rotational symmetry with respect to the [111] axis and we would not expect the observed mirror symmetry for a single crystal. We suggest that the symmetry observed experimentally is caused by 60$^{\circ}$, or equally 180$^{\circ}$, rotational twinning at the bases of the NWs. Most of the CdTe NWs are twin free, except of twinning at the interface to the substrate, as explained in the next section. As in the upper pattern, the spots are a measure for the bulk lattice constant.\\
The electron beam pointing along [1$\overline{1}$0] is deflected during reflection from (11$\overline{2}$) side facets, which show a more complex pattern compared to the atomically flat ($\overline{1}$10) side facets. Instead of continuous lines, small and vertically staggered stripes are visible. One of these stripes is indicated by an arrow in reciprocal space shown left-hand side in Fig.~\ref{RHEED_facets}\,(b).\\
The stripes are located within bands marked with two thin lines in Fig.~\ref{RHEED_facets}\,(b) parallel to the [11$\overline{1}$] direction. Half of the stripes are located between two spots and the position of the remaining stripes is identical to the position of the spots. This interference pattern originates from a regularly stepped surface\,\textsuperscript{\cite{hottier1977}} consisting of (11$\overline{1}$) nanofacets. The step length of 2.2\,nm along [111] is twice the bulk lattice constant in that direction. The deduced cross-sectional view of the atomic positions within the NW and on its (11$\overline{2}$) surface is shown right-hand side in Fig.~\ref{RHEED_facets}\,(b).

\subsection{HgTe epitaxy on CdTe nanowires}
\begin{figure}[h!]
	\begin{center}
		\includegraphics[width=8.5cm]{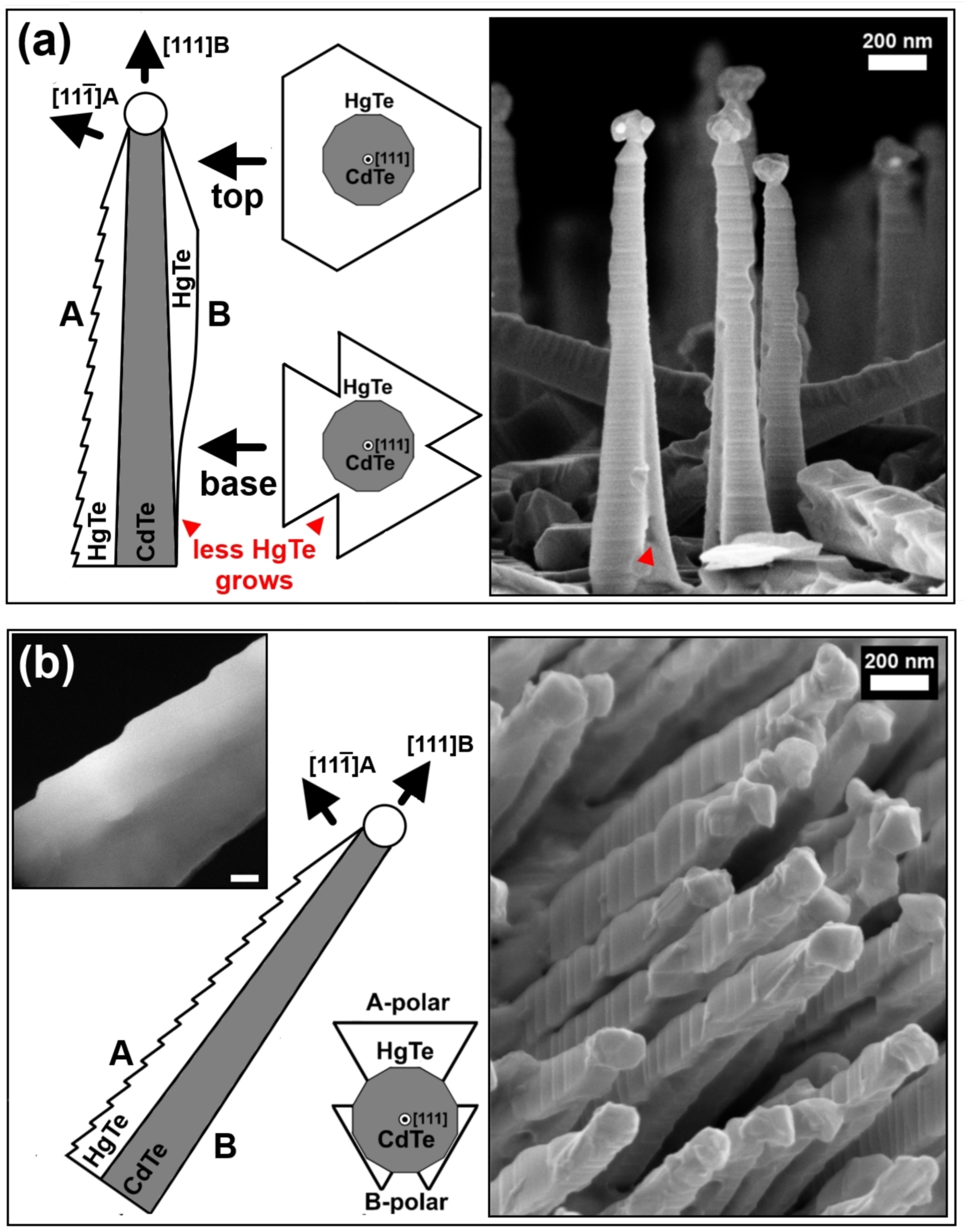}
		\caption[]{Cross-sectional schematic and SEM picture of a sample with a closed HgTe shell around the CdTe NWs is shown in (a). HgTe grown on one side of CdTe NWs are shown in (b) together with the transmission electron micrograph in the upper left, where the scale bar equals 20\,nm.}
		\label{nanowire_shell}
	\end{center}
\end{figure}

This section describes the growth of HgTe shell on the CdTe NWs. HgTe growth occurs at temperatures of about 185 $^{\circ}$C. The sticking of Hg depends strongly on the polarity of the substrate.\textsuperscript{\cite{koestner1988}} We find that Hg flux intensity is a critical parameter due to the multifaceted nature of our NWs. The twelve side facets discussed in the previous section have various polarities: A-polar (group II terminated), non-polar, B-polar (group VI terminated), non-polar etc. Supplying a Hg:Te ratio of 200:1, a continuous HgTe shell grows. The nucleation starts at both polar types of facets and the adjacent non-polar facets are overgrown subsequently. Such CdTe NWs with continuous HgTe shells are shown in Fig.~\ref{nanowire_shell}\,(a).\\
\\
Different growth on polar surfaces lead to the characteristic shell shape. During HgTe overgrowth the stepped and A-polar nanofacets evolve to larger steps resolved by the SEM. The angle between an (11$\overline{1}$)A step and the NWs' [111]B growth direction is determined by the crystal structure. For B-polar facets no pronounced steps are observed in the SEM. Instead, the growth rate of HgTe is reduced gradually on B-side, leading to uncovered areas near the NW bases, as highlighted (red) in Fig.~\ref{nanowire_shell}\,(a). This behavior can be understood from the growth rate dependence on the supplied Hg:Te ratio.\\
We find suppressed growth on the B-polar facets with Hg:Te ratios larger than 250:1, while HgTe still grows stoichiometrically for ratios up to 350:1 on A-polar facets. Suppressing growth on B-polar surfaces can be used to obtain another geometry, shown in Fig.~\ref{nanowire_shell}\,(b). Using a (110) oriented substrate, overgrowth occurs on one side, since such wires are tilted by 35$^\circ$ with respect to the substrate normal. Due to sample-source geometry, the (11$\overline{2}$)A surface facing the effusion cells has a much larger growth rate compared to those A-polar facets, which are oriented more towards the substrate.\\
We have seen that the growth rate dependence on the Hg:Te ratio and the sample-source geometry have an influence on the heterostructure geometry. Evaporated materials have to be multiply absorbed and desorbed from the NW ensemble to reach the bases of long NWs. Impinging Te is more likely incorporated and much less reflected by the NWs than Hg. Thus, the Hg:Te ratio is effectively larger for the NW bases compared to the top ends. This explains the changes in growth rate on B-polar facets along the wire in Fig.~\ref{nanowire_shell}\,(a).\\\\
The differences in growth on the polar facets also reveal the presence of the aforementioned rotational twins in the CdTe wires. The two wires seen in the front of Fig.~\ref{nanowire_shell}(a) are rotational twins. The azimuthal orientation of polar surfaces is constant along each individual wire, thus no further twinning occurs along the wires. We therefore conclude that the origin of the twinning is at the interface to the substrate. The obtained heterostructures have the characteristic shape of a single crystal with faceted side surfaces. We therefore conclude that most of the individual NW heterostructures are single crystalline with a good crystal quality.

\subsection{Charge transport}
\begin{figure}[ht]
	\begin{center}
		\includegraphics[width=8.5cm]{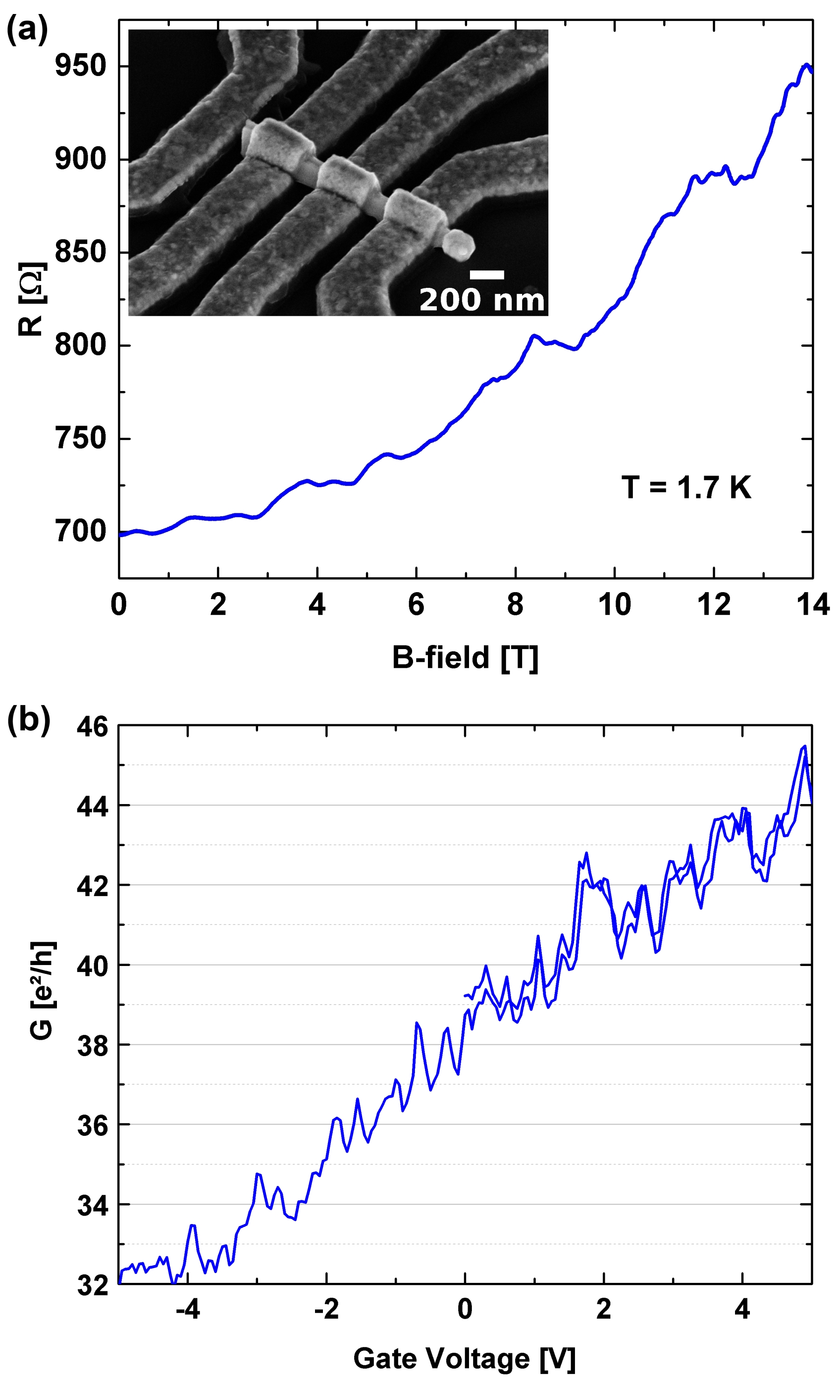}
		\caption[]{(a) Magnetoresistance of a single NW in a perpendicular magnetic field. (b) The effect of a voltage applied to the backgate of another HgTe-covered wire.}
		\label{transport}
	\end{center}
\end{figure}

In addition to the crystallographic analysis presented in the previous parts of the paper we also performed basic transport studies on CdTe-HgTe core-shell structures and uncovered CdTe NW cores.\\\\
The uncovered CdTe wires are not conducting, as we can expect from the 1.4\,eV band gap of this material. In contrast, we find a high conductivity in the structures overgrown with HgTe, even at low temperature (T=1.7\,K). Since these wires are nominally undoped, it is likely that the conductance occurs in the topological surface states of the HgTe capping layer. Figure~\ref{transport}\,(a) shows a magneto-resistance measurement on a single wire sample, as depicted in the inset.
The magnetic field is applied perpendicular to the sample plane and results in a positive and parabolic magnetoresistance consistent with the behavior found in many metallic systems.\textsuperscript{\cite{pippard1989}}
Additional, the resistance shows reproducible fluctuations, which are a manifestation of quantum interference of possible carrier trajectories inside the sample.\\
\\
The conductivity can also be modulated by applying a voltage to the backgate. As seen in Figure~\ref{transport}\,(b), the gate effect leads to a similar reproducible fluctuation of the conductance, pointing to universal conductance fluctuations (UCFs). We furthermore do not observe signs for ballistic transport and therefore conclude, that the mean free path for charge carriers in our quasi one-dimensional HgTe is smaller, but the phase coherence length is larger than the sample size of several hundred nanometers. While HgTe layers show no significant bulk- but surface-transport, the present data can not rule out that this is also true for our NWs. Further experiments, including detailed investigation of gate effects to elucidate these issues are underway.

\section{Summary}
We report on a novel growth method for quasi one-dimensional heterostructures containing the TI material HgTe. The concept of a hollow TI-NW, having a trivial insulator as a core is realized for the first time. The growth of CdTe-HgTe core-shell NWs requires high attention in controlling basic parameters such as substrate temperature and the intensity of supplied material fluxes. We successfully optimized the growth conditions enabling us to grow high quality CdTe NWs, using RHEED for fine tuning of the substrate temperature to achieve straight and long wires. With optimized growth parameters a periodic diffraction pattern allows for the detailed analysis of atomic arrangement on the surfaces and in the bulk. The ability to do so reflects the high crystal quality and ensemble uniformity of our CdTe NWs. The high quality of the cores allows us to grow epitaxial HgTe shells around them.
The wire structures are used for further investigations on the properties of quasi one-dimensional HgTe. So far, we are able to contact a single NW. Bare CdTe cores are insulating, but the HgTe shell conducts and shows UCFs. A full study of the transport properties will be published elsewhere.

\section{Acknowledgements}
This work was supported by the Deutsche Forschungsgemeinschaft in the priority progam ``Topological Insulators: Materials - Fundamental Properties - Devices'' (DFG-SPP 1666).\\
We thank N.\ V.\ Tarakina for the electron transmission micrograph in the inset of Fig.~\ref{nanowire_shell}\,(b).

\bibliography{Ref}

\begin{thebibliography}{19}
\expandafter\ifx\csname natexlab\endcsname\relax\def\natexlab#1{#1}\fi
\expandafter\ifx\csname bibnamefont\endcsname\relax
  \def\bibnamefont#1{#1}\fi
\expandafter\ifx\csname bibfnamefont\endcsname\relax
  \def\bibfnamefont#1{#1}\fi
\expandafter\ifx\csname citenamefont\endcsname\relax
  \def\citenamefont#1{#1}\fi
\expandafter\ifx\csname url\endcsname\relax
  \def\url#1{\texttt{#1}}\fi
\expandafter\ifx\csname urlprefix\endcsname\relax\def\urlprefix{URL }\fi
\providecommand{\bibinfo}[2]{#2}
\providecommand{\eprint}[2][]{\url{#2}}

\bibitem[{\citenamefont{Bernevig et~al.}(2006)\citenamefont{Bernevig, Hughes,
  and Zhang}}]{bernevig2006}
\bibinfo{author}{\bibfnamefont{B.~A.} \bibnamefont{Bernevig}},
  \bibinfo{author}{\bibfnamefont{T.~L.} \bibnamefont{Hughes}},
  \bibnamefont{and} \bibinfo{author}{\bibfnamefont{S.-C.} \bibnamefont{Zhang}},
  \bibinfo{journal}{Science} \textbf{\bibinfo{volume}{314}},
  \bibinfo{pages}{1757} (\bibinfo{year}{2006}).

\bibitem[{\citenamefont{K{\"o}nig et~al.}(2007)\citenamefont{K{\"o}nig,
  Wiedmann, Br{\"u}ne, Roth, Buhmann, Molenkamp, Qi, and Zhang}}]{koenig2007}
\bibinfo{author}{\bibfnamefont{M.}~\bibnamefont{K{\"o}nig}},
  \bibinfo{author}{\bibfnamefont{S.}~\bibnamefont{Wiedmann}},
  \bibinfo{author}{\bibfnamefont{C.}~\bibnamefont{Br{\"u}ne}},
  \bibinfo{author}{\bibfnamefont{A.}~\bibnamefont{Roth}},
  \bibinfo{author}{\bibfnamefont{H.}~\bibnamefont{Buhmann}},
  \bibinfo{author}{\bibfnamefont{L.~W.} \bibnamefont{Molenkamp}},
  \bibinfo{author}{\bibfnamefont{X.-L.} \bibnamefont{Qi}}, \bibnamefont{and}
  \bibinfo{author}{\bibfnamefont{S.-C.} \bibnamefont{Zhang}},
  \bibinfo{journal}{Science} \textbf{\bibinfo{volume}{318}},
  \bibinfo{pages}{766} (\bibinfo{year}{2007}).

\bibitem[{\citenamefont{Fu and Kane}(2007)}]{fu2007}
\bibinfo{author}{\bibfnamefont{L.}~\bibnamefont{Fu}} \bibnamefont{and}
  \bibinfo{author}{\bibfnamefont{C.~L.} \bibnamefont{Kane}},
  \bibinfo{journal}{Phys. Rev. B} \textbf{\bibinfo{volume}{76}},
  \bibinfo{pages}{045302} (\bibinfo{year}{2007}).

\bibitem[{\citenamefont{Br\"une et~al.}(2011)\citenamefont{Br\"une, Liu, Novik,
  Hankiewicz, Buhmann, Chen, Qi, Shen, Zhang, and Molenkamp}}]{bruene2011}
\bibinfo{author}{\bibfnamefont{C.}~\bibnamefont{Br\"une}},
  \bibinfo{author}{\bibfnamefont{C.~X.} \bibnamefont{Liu}},
  \bibinfo{author}{\bibfnamefont{E.~G.} \bibnamefont{Novik}},
  \bibinfo{author}{\bibfnamefont{E.~M.} \bibnamefont{Hankiewicz}},
  \bibinfo{author}{\bibfnamefont{H.}~\bibnamefont{Buhmann}},
  \bibinfo{author}{\bibfnamefont{Y.~L.} \bibnamefont{Chen}},
  \bibinfo{author}{\bibfnamefont{X.~L.} \bibnamefont{Qi}},
  \bibinfo{author}{\bibfnamefont{Z.~X.} \bibnamefont{Shen}},
  \bibinfo{author}{\bibfnamefont{S.~C.} \bibnamefont{Zhang}}, \bibnamefont{and}
  \bibinfo{author}{\bibfnamefont{L.~W.} \bibnamefont{Molenkamp}},
  \bibinfo{journal}{Phys. Rev. Lett.} \textbf{\bibinfo{volume}{106}},
  \bibinfo{pages}{126803} (\bibinfo{year}{2011}).

\bibitem[{\citenamefont{Egger et~al.}(2010)\citenamefont{Egger, Zazunov, and
  Yeyati}}]{egger2010}
\bibinfo{author}{\bibfnamefont{R.}~\bibnamefont{Egger}},
  \bibinfo{author}{\bibfnamefont{A.}~\bibnamefont{Zazunov}}, \bibnamefont{and}
  \bibinfo{author}{\bibfnamefont{A.~L.} \bibnamefont{Yeyati}},
  \bibinfo{journal}{Phys. Rev. Lett.} \textbf{\bibinfo{volume}{105}},
  \bibinfo{pages}{136403} (\bibinfo{year}{2010}).

\bibitem[{\citenamefont{Peng et~al.}(2010)\citenamefont{Peng, Lai, Kong,
  Meister, Chen, Qi, Zhang, Shen, and Cui}}]{peng2010}
\bibinfo{author}{\bibfnamefont{H.}~\bibnamefont{Peng}},
  \bibinfo{author}{\bibfnamefont{K.}~\bibnamefont{Lai}},
  \bibinfo{author}{\bibfnamefont{D.}~\bibnamefont{Kong}},
  \bibinfo{author}{\bibfnamefont{S.}~\bibnamefont{Meister}},
  \bibinfo{author}{\bibfnamefont{Y.}~\bibnamefont{Chen}},
  \bibinfo{author}{\bibfnamefont{X.-L.} \bibnamefont{Qi}},
  \bibinfo{author}{\bibfnamefont{S.-C.} \bibnamefont{Zhang}},
  \bibinfo{author}{\bibfnamefont{Z.-X.} \bibnamefont{Shen}}, \bibnamefont{and}
  \bibinfo{author}{\bibfnamefont{Y.}~\bibnamefont{Cui}}, \bibinfo{journal}{Nat.
  Mater.} \textbf{\bibinfo{volume}{9}}, \bibinfo{pages}{225}
  (\bibinfo{year}{2010}).

\bibitem[{\citenamefont{Hasan and Kane}(2010)}]{hasan2010}
\bibinfo{author}{\bibfnamefont{M.~Z.} \bibnamefont{Hasan}} \bibnamefont{and}
  \bibinfo{author}{\bibfnamefont{C.~L.} \bibnamefont{Kane}},
  \bibinfo{journal}{Rev. Mod. Phys.} \textbf{\bibinfo{volume}{82}},
  \bibinfo{pages}{3045} (\bibinfo{year}{2010}).

\bibitem[{\citenamefont{Mooren}(2010)}]{moore2010}
\bibinfo{author}{\bibfnamefont{J.~E.} \bibnamefont{Mooren}},
  \bibinfo{journal}{Nature} \textbf{\bibinfo{volume}{464}},
  \bibinfo{pages}{194} (\bibinfo{year}{2010}).

\bibitem[{\citenamefont{Xiu et~al.}(2011)\citenamefont{Xiu, He, Wang, Cheng,
  Chang, Huang, Zhou, Jiang, Chen, Zou et~al.}}]{xiu2011}
\bibinfo{author}{\bibfnamefont{F.}~\bibnamefont{Xiu}},
  \bibinfo{author}{\bibfnamefont{L.}~\bibnamefont{He}},
  \bibinfo{author}{\bibfnamefont{y.}~\bibnamefont{Wang}},
  \bibinfo{author}{\bibfnamefont{L.}~\bibnamefont{Cheng}},
  \bibinfo{author}{\bibfnamefont{M.}~\bibnamefont{Chang},
  \bibfnamefont{L.-T.~Lang}},
  \bibinfo{author}{\bibfnamefont{X.}~\bibnamefont{Huang},
  \bibfnamefont{G.~Kou}},
  \bibinfo{author}{\bibfnamefont{Y.}~\bibnamefont{Zhou}},
  \bibinfo{author}{\bibfnamefont{X.}~\bibnamefont{Jiang}},
  \bibinfo{author}{\bibfnamefont{Z.}~\bibnamefont{Chen}},
  \bibinfo{author}{\bibfnamefont{J.}~\bibnamefont{Zou}}, \bibnamefont{et~al.},
  \bibinfo{journal}{Nature Nanotech.} \textbf{\bibinfo{volume}{6}},
  \bibinfo{pages}{216} (\bibinfo{year}{2011}).

\bibitem[{\citenamefont{Hong et~al.}(2014)\citenamefont{Hong, Zhang, Cha, Qi,
  and Cui}}]{hong2014}
\bibinfo{author}{\bibfnamefont{S.~S.} \bibnamefont{Hong}},
  \bibinfo{author}{\bibfnamefont{Y.}~\bibnamefont{Zhang}},
  \bibinfo{author}{\bibfnamefont{J.~J.} \bibnamefont{Cha}},
  \bibinfo{author}{\bibfnamefont{X.-L.} \bibnamefont{Qi}}, \bibnamefont{and}
  \bibinfo{author}{\bibfnamefont{Y.}~\bibnamefont{Cui}}, \bibinfo{journal}{Nano
  Letters} \textbf{\bibinfo{volume}{14}}, \bibinfo{pages}{2815}
  (\bibinfo{year}{2014}).

\bibitem[{\citenamefont{Haakenaasen et~al.}(2008)\citenamefont{Haakenaasen,
  Selvig, Foss, Trosdahl-Iversen, and Taftø}}]{haakennaasen2006}
\bibinfo{author}{\bibfnamefont{R.}~\bibnamefont{Haakenaasen}},
  \bibinfo{author}{\bibfnamefont{E.}~\bibnamefont{Selvig}},
  \bibinfo{author}{\bibfnamefont{S.}~\bibnamefont{Foss}},
  \bibinfo{author}{\bibfnamefont{L.}~\bibnamefont{Trosdahl-Iversen}},
  \bibnamefont{and} \bibinfo{author}{\bibfnamefont{J.}~\bibnamefont{Taftø}},
  \bibinfo{journal}{Applied Physics Letters} \textbf{\bibinfo{volume}{92}},
  \bibinfo{pages}{133108} (\bibinfo{year}{2008}).

\bibitem[{\citenamefont{Neretina et~al.}(2011)\citenamefont{Neretina, Hughes,
  Britten, Sochinskii, Preston, and Mascher}}]{neretina2007}
\bibinfo{author}{\bibfnamefont{S.}~\bibnamefont{Neretina}},
  \bibinfo{author}{\bibfnamefont{R.~A.} \bibnamefont{Hughes}},
  \bibinfo{author}{\bibfnamefont{J.~F.} \bibnamefont{Britten}},
  \bibinfo{author}{\bibfnamefont{N.~V.} \bibnamefont{Sochinskii}},
  \bibinfo{author}{\bibfnamefont{J.~S.} \bibnamefont{Preston}},
  \bibnamefont{and} \bibinfo{author}{\bibfnamefont{P.}~\bibnamefont{Mascher}},
  \bibinfo{journal}{Nanotechnology} \textbf{\bibinfo{volume}{18}},
  \bibinfo{pages}{275301} (\bibinfo{year}{2011}).

\bibitem[{\citenamefont{Huang et~al.}(2014)\citenamefont{Huang, Lu, Chang,
  Banerjee, Hellwarth, and Lu}}]{huang2014}
\bibinfo{author}{\bibfnamefont{L.}~\bibnamefont{Huang}},
  \bibinfo{author}{\bibfnamefont{S.}~\bibnamefont{Lu}},
  \bibinfo{author}{\bibfnamefont{P.}~\bibnamefont{Chang}},
  \bibinfo{author}{\bibfnamefont{K.}~\bibnamefont{Banerjee}},
  \bibinfo{author}{\bibfnamefont{R.}~\bibnamefont{Hellwarth}},
  \bibnamefont{and} \bibinfo{author}{\bibfnamefont{J.~G.} \bibnamefont{Lu}},
  \bibinfo{journal}{Nano Research} \textbf{\bibinfo{volume}{7}},
  \bibinfo{pages}{228} (\bibinfo{year}{2014}).

\bibitem[{\citenamefont{Wojtowicz et~al.}(2008)\citenamefont{Wojtowicz, Janik,
  Zaleszczyk, Sadowski, Karczewski, D{\l}u{\.z}ewski, Kret, Szuszkiewicz,
  Dynowska, Domagala et~al.}}]{wojtowicz2008}
\bibinfo{author}{\bibfnamefont{T.}~\bibnamefont{Wojtowicz}},
  \bibinfo{author}{\bibfnamefont{E.}~\bibnamefont{Janik}},
  \bibinfo{author}{\bibfnamefont{W.}~\bibnamefont{Zaleszczyk}},
  \bibinfo{author}{\bibfnamefont{J.}~\bibnamefont{Sadowski}},
  \bibinfo{author}{\bibfnamefont{G.}~\bibnamefont{Karczewski}},
  \bibinfo{author}{\bibfnamefont{P.}~\bibnamefont{D{\l}u{\.z}ewski}},
  \bibinfo{author}{\bibfnamefont{S.}~\bibnamefont{Kret}},
  \bibinfo{author}{\bibfnamefont{W.}~\bibnamefont{Szuszkiewicz}},
  \bibinfo{author}{\bibfnamefont{E.}~\bibnamefont{Dynowska}},
  \bibinfo{author}{\bibfnamefont{J.}~\bibnamefont{Domagala}},
  \bibnamefont{et~al.}, \bibinfo{journal}{Journal of the Korean Physical
  Society} \textbf{\bibinfo{volume}{53}}, \bibinfo{pages}{3055}
  (\bibinfo{year}{2008}).

\bibitem[{\citenamefont{Janik et~al.}(2007)\citenamefont{Janik,
  D{\l}u{\.z}ewski, Kret, Presz, Kirmse, Neumann, Zaleszczyk, Baczewski,
  Petroutchik, Dynowska et~al.}}]{janik2007}
\bibinfo{author}{\bibfnamefont{E.}~\bibnamefont{Janik}},
  \bibinfo{author}{\bibfnamefont{P.}~\bibnamefont{D{\l}u{\.z}ewski}},
  \bibinfo{author}{\bibfnamefont{S.}~\bibnamefont{Kret}},
  \bibinfo{author}{\bibfnamefont{A.}~\bibnamefont{Presz}},
  \bibinfo{author}{\bibfnamefont{H.}~\bibnamefont{Kirmse}},
  \bibinfo{author}{\bibfnamefont{W.}~\bibnamefont{Neumann}},
  \bibinfo{author}{\bibfnamefont{W.}~\bibnamefont{Zaleszczyk}},
  \bibinfo{author}{\bibfnamefont{L.}~\bibnamefont{Baczewski}},
  \bibinfo{author}{\bibfnamefont{A.}~\bibnamefont{Petroutchik}},
  \bibinfo{author}{\bibfnamefont{E.}~\bibnamefont{Dynowska}},
  \bibnamefont{et~al.}, \bibinfo{journal}{Nanotechnology}
  \textbf{\bibinfo{volume}{18}}, \bibinfo{pages}{475606}
  (\bibinfo{year}{2007}).

\bibitem[{\citenamefont{Tchernycheva et~al.}(2006)\citenamefont{Tchernycheva,
  Harmand, Patriarche, Travers, and Cirlin}}]{tchernycheva2006}
\bibinfo{author}{\bibfnamefont{M.}~\bibnamefont{Tchernycheva}},
  \bibinfo{author}{\bibfnamefont{J.}~\bibnamefont{Harmand}},
  \bibinfo{author}{\bibfnamefont{G.}~\bibnamefont{Patriarche}},
  \bibinfo{author}{\bibfnamefont{L.}~\bibnamefont{Travers}}, \bibnamefont{and}
  \bibinfo{author}{\bibfnamefont{G.~E.} \bibnamefont{Cirlin}},
  \bibinfo{journal}{Nanotechnology} \textbf{\bibinfo{volume}{17}},
  \bibinfo{pages}{4025} (\bibinfo{year}{2006}).

\bibitem[{\citenamefont{Hottier et~al.}(1977)\citenamefont{Hottier, Theeten,
  Masson, and Domange}}]{hottier1977}
\bibinfo{author}{\bibfnamefont{F.}~\bibnamefont{Hottier}},
  \bibinfo{author}{\bibfnamefont{J.}~\bibnamefont{Theeten}},
  \bibinfo{author}{\bibfnamefont{A.}~\bibnamefont{Masson}}, \bibnamefont{and}
  \bibinfo{author}{\bibfnamefont{J.}~\bibnamefont{Domange}},
  \bibinfo{journal}{Surface Science} \textbf{\bibinfo{volume}{65}},
  \bibinfo{pages}{563} (\bibinfo{year}{1977}).

\bibitem[{\citenamefont{Koestner and Schaake}(1988)}]{koestner1988}
\bibinfo{author}{\bibfnamefont{R.}~\bibnamefont{Koestner}} \bibnamefont{and}
  \bibinfo{author}{\bibfnamefont{H.}~\bibnamefont{Schaake}},
  \bibinfo{journal}{Journal of Vacuum Science \& Technology A: Vacuum,
  Surfaces, and Films} \textbf{\bibinfo{volume}{6}}, \bibinfo{pages}{2834}
  (\bibinfo{year}{1988}).

\bibitem[{\citenamefont{Pippard}(1989)}]{pippard1989}
\bibinfo{author}{\bibfnamefont{A.~B.} \bibnamefont{Pippard}},
  \emph{\bibinfo{title}{Magnetoresistance in metals}}, vol.~\bibinfo{volume}{2}
  (\bibinfo{publisher}{Cambridge University Press}, \bibinfo{year}{1989}).

\end{thebibliography}

\end{document}